\documentclass[aps,twocolumn,groupedaddress,showpacs]{revtex4}
\usepackage{epsfig,amsbsy,bm}

\usepackage[colorlinks,linkcolor=black,urlcolor=blue,citecolor=blue]{hyperref}

\newcommand{\eref}[1] {(\ref{#1})}
\newcommand{\Eref}[1] {Eq.~(\ref{#1})}
\newcommand{\Fref}[1] {Fig. \ref{#1}}
\newcommand{\Tref}[1] {Table \ref{#1}}
\newcommand{\be}{\begin{equation}}
\newcommand{\ee}{\end{equation}}
\newcommand{\mc}{\multicolumn}
\newcommand{\ra}{\rangle}
\newcommand{\la}{\langle}

\renewcommand{\bbox}{\boldsymbol }
\newcommand{\cw}{\columnwidth}
\newcommand{\hs}{\hspace*}
\newcommand{\vs}{\vspace*}
\newcommand{\np}{\newpage}
\newcommand{\bt}{\begin{tabular}}
\newcommand{\et}{\end{tabular}}
\newcommand{\bp}{\begin{minipage}}
\newcommand{\ep}{\end{minipage}}

\newcommand{\br}{\begin{eqnarray*}}
\newcommand{\er}{\end{eqnarray*}}
\newcommand{\ba}{\begin{eqnarray}}
\newcommand{\ea}{\end{eqnarray}}
\renewcommand{\r}{{\bbox r}}

\renewcommand{\k}{{\bbox k}} 

\newcommand{\nn}{\nonumber}
\newcommand{\isum}%
{\mathop{\hbox{$\displaystyle\sum\kern-13.2pt\int\kern1.5pt$}}}

\begin{document}

\bibliographystyle{apsrev}

\title
{Time delay in valence shell photoionization of noble gas atoms}

\author{ A. S. Kheifets}
\email{A.Kheifets@anu.edu.au}

\affiliation{Research School of Physical Sciences,
The Australian National University,
Canberra ACT 0200, Australia}

\date{\today}

\begin{abstract}
We use the non-relativistic random phase approximation with exchange
to perform calculations of valence shell photoionization of Ne, Ar, Kr
and Xe from their respective thresholds to photon energy of
200~eV. The energy derivative of the complex phase of the
photoionization matrix elements is converted to the photoelectron
group delay that can be measured in attosecond streaking or two-photon
sideband interference experiments. Comparison with reported time
delay measurements in Ne and Ar at a few selected photon energies is
made. Systematic mapping of time delay across a wide range of photon
energies in several atomic targets allows to highlight important
aspects of fundamental atomic physics that can be probed by attosecond
time delay measurements.
\end{abstract}


\maketitle

\section{Introduction}

Time delay in atomic photoionization has become an active and rapidly
expanding field of research following pioneering experiments on
attosecond streaking \cite{M.Schultze06252010} and two-photon
sideband interference \cite{PhysRevLett.106.143002}. Both techniques use the
XUV pump pulse to ionize the target atom and the IR probe to obtain
the timing information on the photoemission process.
In attosecond streaking, the varying time delay between the the pump
and probe pulses is mapped onto the photoelectron kinetic energy. The
whole valence shell is projected onto a photoelectron kinetic energy
map (the so-called spectrogram) which is then modeled, in the strong
field or Coulomb-Volkov approximations, with the photoionization time
delay being treated as a fitting parameter. This measurement revealed
a relative time delay of $21\pm5$~as between photoemission from the
$2p$ and $2s$ sub-shells in Ne at 106~eV photon energy. The positive sign
of the relative time delay indicates that emission of the
photoelectron from the $2p$ sub-shell is seemingly delayed relative to
that from the $2s$ sub-shell.

In the two-photon interferometric technique, the varying time delay
between the pump and probe pulses is mapped onto the two-photon
sideband (SB) oscillations. The phase of these oscillations depends on
the phase difference of the two neighboring harmonics and the time
delay in atomic photoionization process. The atomic time delay can be
presented as the sum of time delays in the XUV photon absorption and
subsequent IR photon absorption (continuum-continuum or CC transition).
\be
\label{CC}
\tau_A = \tau_W + \tau_{CC}  \ .
\ee
The $ \tau_W$ term represents the Eisenbud-Wigner-Smith time delay (or
Wigner time delay or photoelectron group delay, all these terms will
be used interchangeably in the present context) which is defined as
the energy derivative of the complex phase of the quantum amplitude of
XUV absorption \cite{M.Schultze06252010,PhysRevLett.105.233002}. More
details on the Wigner time delay theory can be found in the review
article \cite{deCarvalho200283}.
The $\tau_{CC}$ term is modeled using the lowest order perturbation
theory and asymptotic forms of the continuum wave functions thus
allowing to obtain the former from an experimental measurement
\cite{Dahlstrom2012}. By reconstructing the oscillations of the SB 22
to 26 of the titanium:sapphire laser at 800~nm,
\citet{PhysRevLett.106.143002} reported the relative time delay
between the photoelectron emission from the $3s$ and $3p$ sub-shells of Ar
in the photon energy range of 34 to 40~eV.
Whether the $3p$ electron was delayed relative to the $3s$ one or vice
versa was found to depend on the photon energy. This measurement was
repeated later by \citet{PhysRevA.85.053424} and the sign of the
relative time delay was reverted with the $3s$ photoelectron being
delayed relative to the $3p$ one near the top end of the photon energy
scale.

This repeated measurement was prompted by observation that the photon
energy of 40~eV fell very close to the Cooper minimum of the $3s$
shell. Photoionization process in this region is driven very strongly
by the many-electron correlation between the $3s$ and $3p$ sub-shells
\cite{PhysRevLett.33.671}.  Such a process cannot be theoretically
described using an independent electron model like the Hartree-Fock
(HF) theory. So the interpretation of the two-photon interferometric
measurement \cite{PhysRevLett.106.143002} based on this theory had to
be re-evaluated. A more adequate model that accounts for inter-shell
correlation in noble gas atoms is the random phase approximation with
exchange (RPAE or, shorter, RPA, both acronyms are used here
interchangeably) \cite{Amusia1972361}.  
However, even after including the RPA corrections, the agreement
between theory and experiment did not improve
\cite{PhysRevA.85.053424}.

Theoretical interpretation of the attosecond streaking measurement of
\citet{M.Schultze06252010} is also not straightforward. The group
delay difference between the $2p$ and $2s$ sub-shells in Ne calculated in
the HF approximation is only 6.2~as \cite{PhysRevLett.105.233002}.
With the added RPA correction of 2.2~as, it accounts for less that a
half of the experimental value of $21\pm5$~as.  More accurate
simulations that accounted for both the XUV and IR fields returned
somewhat larger values of $10.2 \pm 1.3$~as \cite{PhysRevA.84.061404}
and $\sim12$~as \cite{PhysRevA.86.061402}.  These values are still far
too small to match the experimental result.

Even though the streaking IR field is relatively weak, its interplay
with the long-range Coulomb potential of the ionic core (the so-called
Coulomb-laser coupling - CLC) makes an additional contribution to the
streaking time delay  \nocite{PhysRevLett.107.213605,
  PhysRevA.82.043405, 0953-4075-44-8-081001,PhysRevA.85.033401}
\cite{PhysRevLett.107.213605}-\cite{PhysRevA.85.033401}.  Similar to
\Eref{CC}, the streaking time delay can be written as
\be
\label{CLC} 
\tau_s = \tau_W + \tau_{CLC} \ .  
\ee
It was suggested in Ref.~\cite{PhysRevLett.107.213605} that
$\tau_{CLC}$ should also include the effect of the short-range part of
the core potential and hence \Eref{CLC} should be modified to contain
twice the Wigner time delay. This would have resolved the difference
between the theoretical
\cite{M.Schultze06252010,PhysRevLett.105.233002} and experimental
\cite{M.Schultze06252010} time delays.  However, subsequent
investigation on the model two-electron system, that mimicked the
energy levels of the valence shell of Ne, proved that this conjecture
is invalid and that \Eref{CLC} holds. Therefore the controversy
surrounding the experiment \cite{M.Schultze06252010} still remains
unresolved.

In the present paper, we concentrate on the Wigner component $ \tau_W$
which enters the atomic time delay \eref{CC} and the streaking time
delay \eref{CLC}  measured in the attosecond interferometric
and streaking experiments, respectively. The corresponding corrections
to the Wigner time delay, $\tau_{CC}$ and $\tau_{CLC}$ are more or
less universal and can be readily evaluated
\cite{PhysRevA.85.033401,PhysRevA.86.061402}. For two-electron atomic
transitions, like photoionization with excitation and double
photoionization, which are strongly driven by electron correlation,
the streaking time delay \eref{CLC} is further modified by the CLC
effect on the inter-electron interaction
\cite{PhysRevLett.108.163001a}.  These two-electron processes,
however, are outside the scope of the present study.  The target
polarization by the streaking IR field can also be safely ignored as
it should be minimal for tightly bound closed shell atoms.

In the present work, we perform systematic investigation of the Wigner
time delay in a series of noble gas atoms from Ne to Xe across a wide
range of photon energies.  We demonstrate that in heavier noble gases,
beyond Ne, the inter-shell correlation, in the form of direct Coulomb
interaction between atomic electrons assigned to different valence
sub-shells, has a strong effect on photoionization process in general,
and the Wigner time delay, in particular. To account for this direct
inter-electron interaction, we employ the RPA method \cite{A90}. This
method can be viewed as an extension of the HF theory. The latter
accounts for the Coulomb inter-electron interaction only indirectly by
including some part of it in the self-consistent one-electron
potential. On the contrary, the RPA method accounts for a significant
part of the direct inter-electron interaction that results in creation
of pairwise electron-hole excitations. When more complex excitations
of two-electron-two-whole states are important (see e.g. \cite{AK82}),
alternative methods like $R$-matrix \cite{0022-3700-10-18-004} can
provide more accurate results.

We validate our computational technique by making an extensive
comparison between the calculated and experimental valence shell
photoionization cross-sections. Based on this validation, we make
specific predictions for the Wigner time delay and perform further
comparison with available experimental time delay data.  More
generally, we demonstrate that the Wigner time delay contains
important phase information that enables attosecond time delay
measurements to reveal various fundamental aspects of atomic physics.

The paper is organized as follows. In Sec.~\ref{Sec1} we introduce our
computational models for the independent electron descriptions and
that with account for the inter-shell correlations. In Sec.~\ref{Sec2}
we present our numerical results for outer valence $ns$ and $np$
sub-shells in Ne and Ar and $ns$, $np$, $(n-1)d$ sub-shells in Kr and
Xe.  We conclude in Sec.~\ref{Sec3} by revealing the systematic trends
in time delay of noble gases driven by the peculiarities of the
elastic scattering phases and many-electron correlations.

\section{Theoretical model}
\label{Sec1}

\subsubsection{Independent electron HF model}

We adopt the photoionization formalism as outlined in the monograph
\cite{A90}. We evaluate the one-photon dipole matrix element
$\la \psi^{(-)}_\k|\hat z| \phi_i\ra$
of the transition from a bound state $i$ to an incoming continuous
wave with the given photoelectron momentum $\k$.  The magnitude of the
momentum is restricted by the energy conservation
$ E\equiv k^2/2 = \omega+\varepsilon_i \ , $
where $\omega$ is the photon energy. The atomic units are used
throughout the paper with $e=m=\hbar=1$ and the atomic unit of time is
approximately equal to 24.2~as.

We split the radial and angular dependence in the initial state
$ \phi_i(\r) = Y_{l_im_i}(\hat\r)R_{n_il_i}(r) $
and use the partial wave expansion in the final state
\begin{eqnarray}
\label{partial} \psi^{(-)}_{\k}(\r) = 
\frac{(2\pi)^{3/2}}{k^{1/2}}\hs{-1mm}
\displaystyle\sum_{lm} i^l
e^{-i\delta_l(E)}Y^*_{lm}(\hat{\k})Y_{lm}(\hat\r)R_{El}(r) \ ,
\end{eqnarray}
where the radial orbitals are normalized to energy
$
\la El\|E'l\ra = \delta(E-E')
$
and have the asymptotics at infinity
$$
P_{El}(r)\Big|_{r\to\infty} 
=\sqrt{2\over\pi k}\,{1\over r}\,
\sin(kr-{l\pi\over2}+\delta_l) 
\ .
$$
We align the quantization axis $z$ with the polarization axis of light
and write the dipole operator in the length gauge as
$\hat z=\sqrt{4\pi/3}\, rY_{10}(\hat \r) \ .$
We perform the spherical integration to arrive to the following
expression:
\ba
\label{dipole} 
\la \psi^{(-)}_\k|\hat z| \phi_i\ra 
&=&  \frac{(2\pi)^{3/2}}{k^{1/2}}
\sum_{l=l_i\pm1\atop m=m_i}
e^{i\delta_l(E)}i^{-l}Y_{lm}(\hat \k)\, \\&&\hs{5mm}\times
\left(\begin{array}{rrr} l&1&l_i\\ m&0&m_i\\
\end{array}\right) \la El \|\,r\,\|n_il_i\ra 
\nn \ea
Here the reduced dipole matrix element, stripped of all the angular
momentum projections, is defined as
\be
\label{reduced} 
\la El \|\,r\,\|n_il_i\ra = \hat l \hat l_i \left(
\begin{array}{rrr} l&1&l_i\\ 0&0&0\\
\end{array} \right)\\ \int r^2dr \, R_{El}(r)\,r\,R_{n_il_i}(r) \ ,
\ee
where $\hat l = \sqrt{2l+1}$.
The partial photoionization cross section for the transition from an
occupied state $n_il_i$ to the photoelectron continuum state $El$ is
calculated as
\be
\label{CS-HF} \sigma_{n_il_i\to El}(\omega) = 
\frac43 \pi^2\alpha a_0^2\omega
\left| \la E l\,\|\,r\,\|n_il_i \ra \right|^2 \ .  
\ee 
Here $\alpha$ is the fine structure constant and $a_0$ is the Bohr
radius.

The basis of occupied atomic states $\|n_il_i\ra$ is defined by the
self-consistent HF method and calculated using the computer code
\cite{CCR76}. The continuum electron orbitals $\la El \|$ are defined
within the frozen-core HF approximation and evaluated using the
computer code \cite{CCR79}. These states are found in the combined
field of the nucleus and the HF potential of the frozen electron
core. So the photoelectron scattering phase $\delta_l(E)$ delivered by
this method contains both the long-range Coulomb and the short-range
Hartree-Fock components.

We note that the reduced matrix element \eref{reduced} is real and
thus the complex phase of the dipole matrix element \eref{dipole} is
defined by the scattering phases $\delta_{l_i\pm1}(E)$. According to
the Fano's propensity rule \cite{PhysRevA.32.617}, the dipole
transition with the increased momentum $l= l_i+1$ is usually dominant.
In such a situation, the photoemission group delay is approximately
given by $\tau_W = d\delta_l/dE$.

\subsubsection{Inter-shell correlation}

To include inter-shell correlation effects, we employ the RPA model
\cite{A90}. In this approximation, the reduced dipole matrix element
\eref{reduced} is replaced by its correlated counterpart $\la
El\|D\|n_il_i \rangle$ which accounts for correlation between
different valence sub-shells. This correlated matrix element is found
as a solution of the system of the integral equations:
\begin{eqnarray}
\label{RPA} \la El\|D\|n_il_i \ra &=& \la El\|\,r\,\|n_il_i \ra 
+  \frac13 \lim_{ \varepsilon \to 0^+}
\isum_{\scriptstyle n'l'\atop \scriptstyle n_jl_j } dE'\\
&&\hs{-2cm}
\Bigg[ 
{\la E' l'\|D\|n_jl_j \ra 
\la n_jl_jEl\|V\| E' l' n_il_i \ra 
\over 
\omega - E'+\epsilon_{n_jl_j} +i\varepsilon } \nn 
\\ &&\hs{-0.3cm} + 
{ \la n_jl_j \|D\|E' l' \ra 
\la p l' El\|V\|n_jl_jn_il_i \ra 
\over 
\omega + E'-\epsilon_{n_jl_j} } \Bigg], \nn
\end{eqnarray}
Here the combined sum/integral sign incorporates both the summation
over the discrete excited states $n'l'$ with the energy
$\epsilon_{n'l'}$ and the integration over the continuum $dE'$ from
the threshold to infinity.  The Coulomb matrix contains both the
direct and the exchange parts $V=2U-W$. That explains the term
exchange in the name RPA(E). The direct Coulomb matrix is expressed as
\ba
\label{direct}
\la n_jl_jEl\|U\| E' l' n_il_i
&=&
\hat l  \hat l' \, \hat l_i \,\hat l_j
\left(\begin{array}{rrr}
l&1&l_i\\
0&0&0\\
\end{array}\right)
\left(\begin{array}{rrr}
l'&1&l_j\\
0&0&0\\
\end{array}\right)
\nn \\&\times&
R^{(1)}_{l,l',l_i,l_j}(E,E',n_i,n_j) \ ,
\ea
where $R^{(1)}$ is  a Slater integral \cite{A90}.
In the exchange matrix, the electron $ El$ and the hole
$n_jl_j$ states are swapped. 

\begin{figure}[htbp] \centering
  \includegraphics[width=8.0cm] {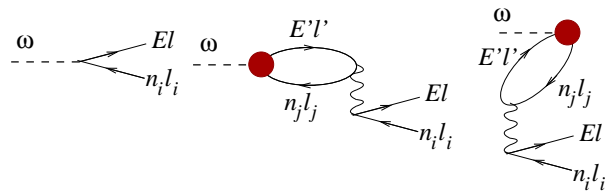} \bigskip
\caption{ Graphical representation of the RPA equations \eref{RPA}.
Left: non-correlated dipole matrix element.  Center: time-forward
process.  Right: time-reverse process.}
\label{Feynman}
\end{figure}

The RPA equations are represented graphically in \Fref{Feynman}.  Here
the straight line with an arrow to the left or right represents
electron (continuum) or hole (bound) states, respectively. The wavy
line exhibits the Coulomb interaction.  The dashed line is used to
display a photon of the frequency $\omega$.  The shaded circle is used
to represent the correlated dipole matrix element whereas the bare
matrix element is exhibited by a three-pronged vertex.  The Coulomb
interaction matrices
$\la n_jl_jEl\|V\|E'l' n_il_i \ra $
and
$\la E'l' El\|V\|n_il_in_jl_j \ra $
describe the so-called time-forward and time-reverse correlation
processes which are exhibited by the second and third diagrams (from
left to right). In the time-forward process, the photon absorption is
followed by the inter-electron interaction in the form of creation of
the virtual electron-hole pair in the neighbouring sub-shell. In the
time-reverse process, the virtual electron-hole pair is created before
the photon absorption takes place. Because the time-forward process is
real in a sense that it conserves the energy of the system, while the
time-backward process is virtual, the time-forward process makes stronger
contribution to the photoionization process.  However, for
the completeness and gauge invariance of the theory, both processes
should be taken into account.

We solve the system of integral equations \eref{RPA} using a slightly
modified version of the computer code \cite{AC97}. While in the
original code correlations between two sub-shells could only be taken
in account, the present version  allows to include an
arbitrary number of sub-shells.  The energy integration in the
time-forward term of \Eref{RPA} (second line) contains a pole and the
RPA matrix element acquires an imaginary part and therefore an extra
phase
$\arg \la El\|D\|n_il_i \ra \ .$
However, this phase does not enter the partial photoionization cross
section $n_il_i\to El$ which is obtained from the squared matrix
element.
To get access to the phase information, one has to evaluate the
angular asymmetry parameter $\beta$  which contains the phase difference
between the two photoionization channels $l=l_i\pm1$ when $l_i\ne0$
\cite{A90}. The photoelectron group delay, which is the energy
derivative of the phase of the complex photoionization amplitude,
gives an alternative access to the phase information. It is evaluated
as
\be 
\label{delay}
\tau = {d\over dE} \arg f(E)\equiv
{\rm Im} \Big[ f'(E)/f(E) \Big] \ .
\ee
Here the photoionization amplitude $f(E)$ is given the partial wave
expansion
\ba
\label{amplitude}
f(E)&\propto&
\sum_{l=l_i\pm1}
e^{i\delta_l}i^{-l}
Y_{lm}(\hat \k)\,
(-1)^m
\left(\begin{array}{rrr}
l&1&l_i\\
-m&0&m_i\\
\end{array}\right)
\nn\\&&\hs{2cm}\times \ 
 \la El\|D\|n_il_i \ra
\ea
The amplitude $f(E)$ is evaluated in the forward direction
$\k\|\hat z$, which is usually the case in the attosecond time delay
measurements. In this case,
$
Y_{lm}(\hat \k\|\hat z)=
\hat l (4\pi)^{-1/2}
\delta_{m0} \ 
$
and hence $m_i=0$ also.
It has to be noted that the phase of the amplitude \eref{amplitude}
contains the contribution of the HF phases $\delta_l$ in both
photoionization channels $l=l_i\pm1$ as well as the RPA correction due
to the imaginary part of the RPA dipole matrix element $\la
El\|D\|n_il_i\ra$. Thus the associated group  delay is labeled HF+RPA
when the numerical results are presented in the following section.

\section{Numerical results}
\label{Sec2}

\subsection{Neon $2s$ and $2p$ sub-shells}

\begin{figure}[h] \vs{0.9\cw} \epsfxsize=0.8\cw

\epsffile{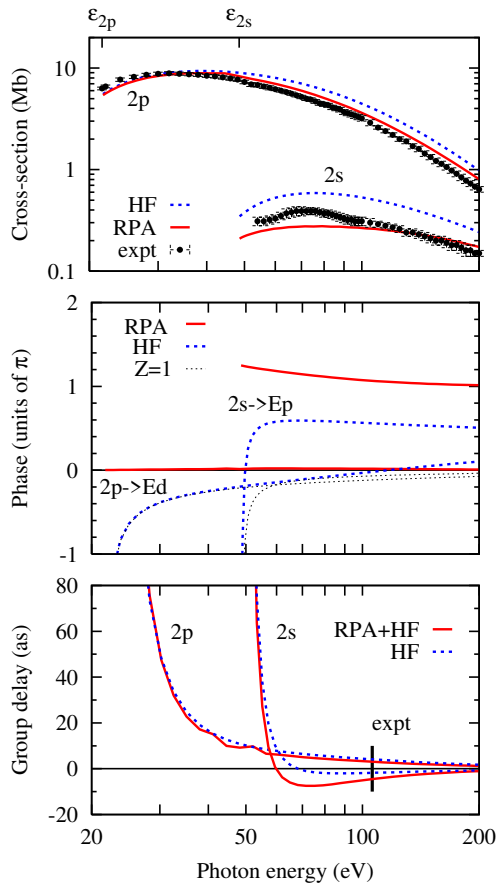}
\caption{(Color online) Top: the partial photoionization
cross-sections of the $2s$ and $2p$ sub-shells of Ne. The HF and RPA
calculations are shown by the dashed (blue) and solid (red) lines,
respectively. The recommended experimental data by
\protect\citet{Bizau1995205} are displayed with error bars. Middle:
elastic scattering phases in the field of the Ne$^+$ ion for the
$2s\to Ep$ and the dominant $2p\to Ed$ channels (dotted blue line) and
the RPA phases (solid red line).  The thin dotted line visualizes the
Coulomb phase with $Z=1$. Bottom: the phase derivatives are converted
to the units of the group delay. The vertical bar at the photon energy
of 106~eV visualizes the relative time delay between the $2p$ and $2s$
sub-shells of 21~as as measured by \protect\citet{M.Schultze06252010}
\label{FigNe}}
\end{figure}

On the top panel of \Fref{FigNe} we present the partial
photoionization cross-sections of valence shell photoionization of Ne.
The HF cross-sections  are shown by the dashed (blue)
lines and the RPA cross-sections  are exhibited by the
solid (red) line. The recommended experimental data by
\protect\citet{Bizau1995205} are displayed with error bars.  In the
RPA calculation, we substitute the HF bound state energies with the
the experimental ionization thresholds $\epsilon_{2p_{3/2}} =
21.56$~eV and $\epsilon_{2s}=48.47$~eV \protect\cite{NIST-ASD} which
are indicated on the upper boundary of the panel. We see that account
for the RPA correlation between the $2s$ and $2p$ sub-shells improves the
calculated cross-sections and makes then closer to the experimental
data.

We note that even though agreement between theory and experiment is
improved in the RPA model, there is a visible difference between
the calculated and measured cross-sections, especially for the $2s$
sub-shell.  This difference may arise from the fact that not all the
many-electron correlations are accounted for by the RPA model which
includes pairwise electron-hole virtual excitations. Other processes
like admixture of the two-hole-one-electron states to the pure
one-hole state in the singly charged ion are not included in the RPA
model. This admixture is responsible for the shift in atomic
ionization potentials relative to the corresponding HF binding
energies as well as appearance of the satellite lines in the
photoionization spectra \cite{AK84a}.  These effects cannot be
accounted for {\em ab initio} in the RPA model.  Phenomenologically,
they are partly compensated by using the experimental ionization potentials
instead of the HF energies $\varepsilon_i$ in the RPA equations \eref{RPA}.

On the middle panel, we show the elastic scattering phases in the
field of the Ne$^+$ ion for the $2s\to Ep$ and the dominant $2p\to Ed$
channels. The HF phases $\delta_p(E)$ and $\delta_d(E)$ are plotted
with the dashed (blue) line. The RPA phases $\arg \la kp\|D\|2s\ra$
and $\arg \la kd\|D\|2p\ra$ are displayed with the solid (red) line.
The thin dotted line visualizes the Coulomb phase
$\sigma_l(E) = \arg \Gamma\left(1+l-i{Z_{\rm eff}/ \sqrt{2E}}\right) $
with the effective charge $Z_{\rm eff}=1$. This phase shows the
contribution of the long-range Coulomb potential to the HF phase
$\delta_l(E)$ which is strongly dominant at small kinetic energies of
the photoelectron.  The phase shift due to the short range potential,
i.e. the difference of the total phase and the Coulomb phase, is
related to the quantum defect according to the Levinson-Seaton theorem
$ \delta_l(k\to0)-\sigma_l(k\to0) = \mu_l(\infty)\pi \ $
\cite{PhysRevA.52.3824}. For a neutral target, the scattering phase at
zero energy is related to the number of the bound target states $N_l$
by the Levinson's theorem $\delta_l(k\to0) = N_l\pi$.  In the absence
of the Coulomb potential, the $2s\to Ep$ phase would tend to one unit
of $\pi$ at $k\to0$ as there is one occupied $np$ sub-shell in the
Ne$^+$ ion with $n=2$. 
With the Coulomb potential taken into account,
$ \delta_{2s\to kp}(k\to0)-\sigma_{2s\to kp}(k\to0) = 0.88\pi \ $
where $\mu_{l=1} = 0.88$ is the quantum defect calculated from fitting
the $np$ orbital energies in the Ne$^+$ ion $\epsilon_{np}\propto
-(n-\mu_{l=1})^2$ for $n>2$.  As the Coulomb phase tends to zero
rapidly away from the threshold, the HF phase stays rather flat at the
value determined by the corresponding quantum defect. We may associate
this behavior with the Levinson theorem, even though this theorem is
strictly valid only at $k\to0$.
Similarly, the the $2p\to Ed$ phase would tend to zero as there are no
occupied $d$-shells left behind.  The Coulomb logarithmic singularity
changes this behaviour radically and sends the scattering phases to
large negative values near the threshold. It has to be noted that when
the Coulomb behavior of the phases and associated group delays becomes
dominant at low photoelectron energies, the measurement-induced
components in the experiments, i.e.  the Coulomb-laser coupling in
attosecond streaking \cite{PhysRevA.82.043405, 0953-4075-44-8-081001}
or the continuum-continuum contribution in interferometric two-photon
measurements \cite{PhysRevLett.106.143002,
  PhysRevA.85.053424,Dahlstrom2012}, will be large and the group delay
as presented in this paper can only be accessed if those corrections
are properly accounted for.

The RPA phase in the $2p\to Ed$ channel is hardly distinguishable from
zero. This observation is consistent with a very small change that the
RPA correction causes to the partial photoionization cross-section
shown on the top panel. The RPA phase in the $2s\to Ep$ channel is
large but rather flat and changes slowly with the photon energy. This
is consistent with the $2s$ partial photoionization cross-section
which is affected by by the inter-shell correlation with $2p$ across
the whole range of the studied photon energies.

The bottom panel of \Fref{FigNe} displays the photoelectron group
delay calculated as the energy derivative of the phase of the
photoionization matrix element. The HF group delay in the dominant
photoionization channel is calculated as $\tau^{HF}_W({\rm as})
=k^{-1}d\delta_l/ dk \times 24.2$. Here $E=k^2/2$ is the photoelectron
energy in atomic units and one unit of time is equal approximately to
24.2~as. In the existing code, the continuous electron orbitals
are calculated on the regular momentum grid and numerical
differentiation over the momentum, rather than energy, is easier to
implement. The fine grid of 0.05~au of photoelectron momentum is
sufficient for an accurate numerical differentiation.  Similarly, the
combined RPA+HF time delay is calculated as
$ \tau^{RPA+HF}_W({\rm as}) = {\rm Im} \Big[ k^{-1}f'(k)/f(k) \Big]
\times 24.2 \ .  $
Here the photoionization amplitude \eref{amplitude} is evaluated in
the $z$-axis direction.  We see that the HF time delay in the dominant
$2p\to Ed$ channel accounts for almost the whole time delay in
photoemission from the $2p$ sub-shell. There is some oscillation
visible due to the autoionizing resonances near the $2s$ threshold
which is absent in the HF approximation. Overall, the $2p$ time delay
is always positive and rapidly decreasing function of the photon
energy. This is explained by the monotonously decreasing HF phase in
the $d$-partial wave which is driven by the Coulomb logarithmic
singularity. The situation is different in the $2s\to Ep$ channel.
Here the HF phase crosses over from the Coulomb behavior at low
photoelectron kinetic energy to the Levinson behavior at larger
energies. In result, the phase derivative and, consequently, the time
delay change their sign from positive and negative towards the larger
photon energies. The RPA correction to the time delay is always
negative. Hence the photoemission from the $2s$ sub-shell seems to be
ahead of that of the $2p$ sub-shell at around 100~eV photon energy
mark where the measurement of \citet{M.Schultze06252010} was taken
(shown as a vertical bar in the figure). According to \eref{CLC}, to
make a comparison of the present calculation with the experiment, we
have to add to the Wigner time delay difference between the $2p$ and
$2s$ sub-shells $\Delta\tau_W=8.4$~as with the difference between the
corresponding CLC corrections $\Delta\tau_{CLC}=3.5$~as
\cite{PhysRevA.85.033401}. The resulting time delay difference
$\Delta\tau_{s}=11.9$~as which is very similar to that reported in
\cite{PhysRevA.86.061402} by only half of the experimental value of
$21\pm5$~as.

\subsubsection{Argon $3s$ and $3p$ sub-shells}

\begin{figure}[h] \vs{0.9\cw} \epsfxsize=0.8\cw
\epsffile{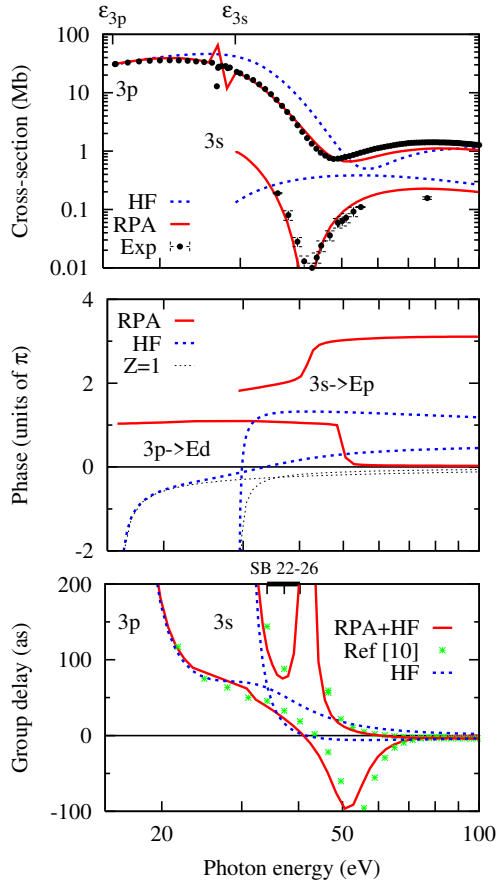}
\caption{(Color online) Top: the partial photoionization
cross-sections of the $3s$ and $3p$ sub-shells of Ar. The HF and RPA
calculations are shown by the dashed (blue) and solid (red) lines,
respectively. The experimental data for $3s$
\protect\cite{PhysRevA.47.3888}  and for $3s+3p$
\protect\cite{Samson2002265}  are displayed with error
bars.  Middle: elastic scattering phases in the field of the Ar$^+$
ion for the $3s\to Ep$ and the dominant $3p\to Ed$ channels (dotted
blue line) and the RPA phases (solid red line). Bottom: the phase
derivatives are converted to the units of the group delay. The green
asterisks display the calculation \cite{PhysRevA.86.061402} corrected
for experimental ionization thresholds.
\label{FigAr}}
\end{figure}

An analogous set of data for Ar $3s$ and $3p$ sub-shells is shown in
\Fref{FigAr}. On the top panel we make a comparison of the HF ( dashed
blue line) and the RPA (solid red line) partial photoionization
cross-sections with the experimental data by \citet{PhysRevA.47.3888}
for $3s$ sub-shell and by \citet{Samson2002265} for the sum of $3s$ and
$3p$ sub-shells.  The experimental ionization thresholds $\rm
\epsilon_{3p_{3/2}} = 15.76$~eV and $\epsilon_{3s}=29.24$~eV
\protect\cite{NIST-ASD} are indicated on the upper boundary of the
panel. These partial photoionization cross-sections are qualitatively
different from those of Ne shown in \Fref{FigNe}. Firstly, the $3p$
cross-section in Ar displays the Cooper minimum whereas the nodeless
$2p$ orbital does not \cite{RevModPhys.40.441}. Second, the
inter-shell correlation changes completely the $3s$ cross-section
which also display a deep Cooper-like minimum at a slightly smaller
photon energy. The RPA calculation reproduces these features in fair
agreement with the experiment.


The HF phases in Ar behave similarly to the analogous case of Ne
except that the $3s\to Ep$ phase would tend to $2\pi$ in the absence
of the Coulomb singularity as there are two occupied $np$-shells in
the Ar$^+$ ion.
With the Coulomb potential taken into account,
$ \delta_{3s\to kp}(k\to0)-\sigma_{3s\to kp}(k\to0) = 1.73\pi \ $
where the corresponding value of the quantum defect in Ar$^+$ is
$\mu_{l=1}=1.73$.  The RPA phases in Ar are very different from Ne.
When the cross-section goes through the Cooper minimum, the
corresponding phase makes a jump of $\pi$ in the $3s\to Ep$ amplitude,
and $-\pi$ in the $3p\to Ed$ amplitude. This jump is easy to
understand. If the amplitude was real and had a node, it would simply
change its sign which would amount to adding a phase factor of $\pi$
in the complex number representation. Incidentally, this jump was
investigated in an earlier model calculation
\cite{PhysRevLett.105.073001} which established validity of the
attosecond streaking technique for the phase measurements.

This jump of $\pi$ has a dramatic effect on the time delay which is
shown on the bottom panel of \Fref{FigAr}. It drives the time delay in
the $3s$ sub-shell to very larger numbers in several hundreds of
attoseconds. The situation is less dramatic for the $3p$ sub-shell. Here
the normally weak transition $3p\to Es$ takes over near the Cooper
minimum of the strong $3p\to Ed$ transition and the resulting time
delay does not go below $-100$~as. We note that there is a strong
variation of phase near the autoionization resonances in the $3p$
photoionization which is seen on the top panel of \Fref{FigAr}. We do
not show this variation in the phase and time delay plots for clarity
of presentation. Anyway, this resonances are far too narrow to be
detected in time delay measurements at present energy resolution.

One can compare significant time delay near the Cooper minimum with
the delay time in Breit-Wigner resonant  scattering $t_d=2/\Gamma$
with $\Gamma$ being the resonant width at half maximum of the cross-section
\cite{Newton1982}.   In the case of the Cooper minimum in Ar, which is
roughly 0.5~a.u. of energy wide, the time delay is expected to be 4
atomic units of time which equates to about 100~as. Of course, this is
a very rough estimate and the actual time delay is not constant but
varies across the Cooper minimum. The steepness of this variation
can only be estimated from an accurate numerical calculation.

On the upper boundary of the bottom panel, we indicate the 
photon energies corresponding to the SB 22 to 26 of the
titanium:sapphire laser at 800~nm used in the two-photon
interferometric experiments
\cite{PhysRevLett.106.143002,PhysRevA.85.053424} We see that at this
photon energy range, the RPA correction changes completely the sign of
the relative $3p/3s$ time delay. In the HF approximation, the $3p$
photoemission is delayed more that the $3s$ ones. The inter-shell
correlation changes this ordering completely. With the RPA correction,
it is the $3s$ that is delayed more than the $3p$. This is an
important, strong and qualitative result which is related to the
Cooper minima in the corresponding partial photoionization
cross-sections. This result is confirmed by an alternative
time-independent calculation by \citet{PhysRevA.86.061402} with a
similar account for many-electron correlations as in RPA. As compared
to the original calculation presented in \cite{PhysRevA.86.061402},
the group delay data shown on the bottom panel of \Fref{FigAr} are
corrected for the experimental ionization potentials
\cite{Marcus2013}. 
Without this correction, the HF ionization potential of the $3s$
sub-shell $\varepsilon_{3s} = 34.7$~eV makes the SB 22 inaccessible.

A strong modification of the relative time delay between the $3p$
and $3s$ sub-shells in Ar is more clearly seen in \Tref{Tab1} where we
present the time delay difference $\tau_{3s}-\tau_{3p}$ in the HF and
RPA approximations and compare it with the experimental data of
\citet{PhysRevA.85.053424}. Even a fairly large uncertainty of $\pm
50$~as cannot reconcile the experimental data with neither of the
calculations. In the same table, we present results 
of a multi-configurational Hartree-Fock (MCHF) close-coupling
calculation \cite{PhysRevA.87.023420}. In this calculation, the
Cooper minimum was displaced to significantly larger photon energies
which were not probed experimentally. Hence the time delay difference
at the SB~26 was not affected by this minimum as strongly as in the
present RPA calculation. 

\begin{table} \bt{r rrr} 
\mc{1}{c}{SB} & \mc{1}{c}{22} & \mc{1}{c}{24} &\mc{1}{c}{26}\\ 
$\omega$ (eV) & 34.1 & 37.2 & 40.3 \\ \hline\hline
$\tau_W^{3s}-\tau_W^{3p}$ (as)& \\ 
HF& 3 & -36 & -38\\
RPA& 76 & 53 & 215\\ 
MCHF& 45  & 10 & -5\\
Expt &70 &-30 & 50\\ 
\et
\caption{Relative time delay between the photoemission from the $3s$
  and $3p$ sub-shells $\Delta\tau_W=\tau_W^{3s}-\tau_W^{3p}$ in Ar 
at three fixed photon
  energies corresponding to the SB 22 to 26 in the experiment of
  \protect\citet{PhysRevA.85.053424}. The experimental uncertainty is
  $\pm 50$~as. The MCHF calculation \cite{PhysRevA.87.023420} has  
typical error bars of $\pm40$~as due to pseudo-resonance structure.
\label{Tab1}}
\end{table}

\subsubsection{Krypton $4p$, $4s$ and $3d$ sub-shells}

\begin{figure}[h] \vs{0.9\cw} \epsfxsize=0.8\cw
\epsffile{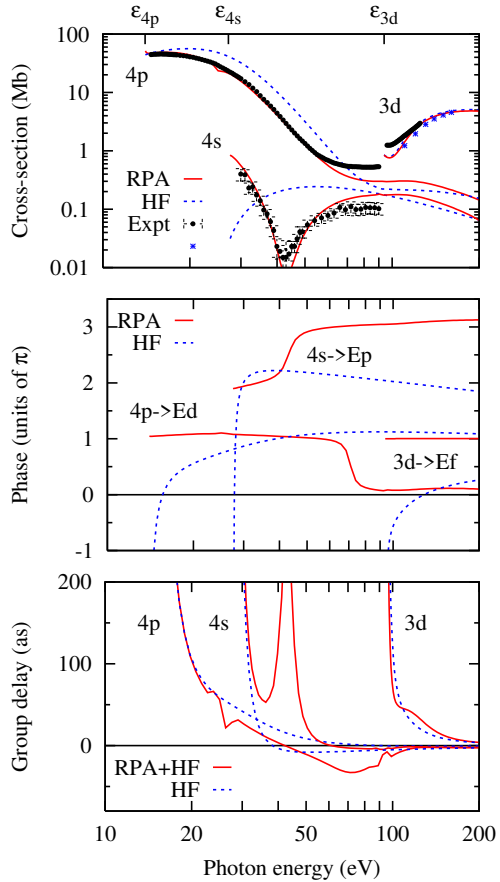}
\caption{(Color online) Top: the partial photoionization
cross-sections of Kr. The HF and RPA calculations are shown by the
dashed (blue) and solid (red for $4s, 4p$ and green for $3d$)) lines,
respectively.
The experimental data by \protect\citet{0953-4075-27-8-010} for $4s$
and by \protect\citet{Samson2002265} for $4p+3d$ are displayed with
error bars. The data from \protect\citet{PhysRevA.36.3449} for $3d$
are displayed with asterisks.
Middle: elastic scattering phases in the field of the Ar$^+$ ion for
the $4s\to Ep$ and the dominant $4p\to Ed$ channels (dotted blue line)
and the RPA phases (solid red line).  Bottom: the phase derivatives
are converted to the units of the group delay.
\label{FigKr}}
\end{figure}

Our results for the $4p$, $4s$ and $3d$ photoionization of Kr are
displayed in \Fref{FigKr}.  On the top panel we make a comparison of
the HF ( dashed blue line) and the RPA (solid red line) partial
photoionization cross-sections with the experimental data of
\citet{0953-4075-27-8-010} for $4s$ and of \citet{Samson2002265} for
$4p+3d$ (error bars). The data from \citet{PhysRevA.36.3449} for $3d$
are displayed with asterisks. The experimental ionization thresholds
$\rm \epsilon_{4p_{3/2}} = 14.00$~eV, $\epsilon_{4s}=27.51$~eV
\cite{NIST-ASD} and $\epsilon_{3d_{5/2}} = 93.83$~eV
\cite{0953-4075-25-18-006} are indicated on the upper boundary of the
panel. The $4p$ and $4s$ cross-sections in Kr behave similarly to the
$3p$ and $3s$ cross-sections in Ar (see the top panel of
\Fref{FigAr}). The $4p\to Ed$ cross-section goes through its Cooper
minimum which is offset somewhat by the weaker $4p\to Es$ channel. So
the total $4p$ cross-section displays a shoulder rather than a true
minimum. The $4s$ cross-section is driven strongly by its inter-shell
correlation with $4p$ to a very deep minimum which is missed
completely in the HF approximation. The $3d$ cross-section from its
threshold displays a strong maximum associated with its shape
resonance. This resonance is known to be due to electron correlation
within a single shell \cite{0034-4885-55-9-003} and indeed the $3d$
photoionization cross-section is well described by the HF
approximation.


The HF phases in Kr (middle panel of \Fref{FigKr}) behave similarly to
the analogous cases of Ne and Ar except that the $4s\to Ep$ phase
would tend to $3\pi$ and the $4p\to Ed$ phase would tend to $\pi$ in
the absence of the Coulomb potential. With this potential, the HF
phases are determined by the corresponding quantum defect values
$\mu_{l=1}=2.67$ and $\mu_{l=2}=1.04$.
The RPA phases in Kr are also similar to Ar. Every time the
cross-section goes through the Cooper minimum, the corresponding
phase makes a jump of $\pi$: upwards in the $4s\to Ep$ amplitude and
downwards in the $4p\to Ed$ amplitude. The RPA phase in the $3d\to Ef$
transition is rather stationary.

This behavior of the phases translates into the corresponding time
delays plotted on the bottom panel of \Fref{FigKr}. The RPA time delay
in $4p$ sub-shell is not dramatically different from the HF
calculation. Even though the dominant $4p\to Ed$ transition displays a
Cooper minimum, it is offset by the weak $4p\to Es$ transition and
is not as prominent in the total $4p$ cross-section as in the $3p$
cross-section of Ar. There are some variation of the time delay near
the autoionizing resonances close to the $4s$ threshold which are seen
in the RPA calculation but not in HF one. The time delay in the $3d$
sub-shell is almost entirely due to intra-shell effects and the HF and RPA
results are very close. The situation is very different in the $4s$
sub-shell where the time delay is strongly affected by the inter-shell
correlation with the $4p$ sub-shell and reaches 300~as in its
peak. Similarly to Ar, there is a complete reversal of the relative
time delay between the $4p$ and $4s$ sub-shells in the RPA calculation in
comparison with the HF one.

\subsubsection{Xenon $5p$, $5s$ and $4d$ sub-shells}

\begin{figure}[h] \vs{0.85\cw} \epsfxsize=0.8\cw
\epsffile{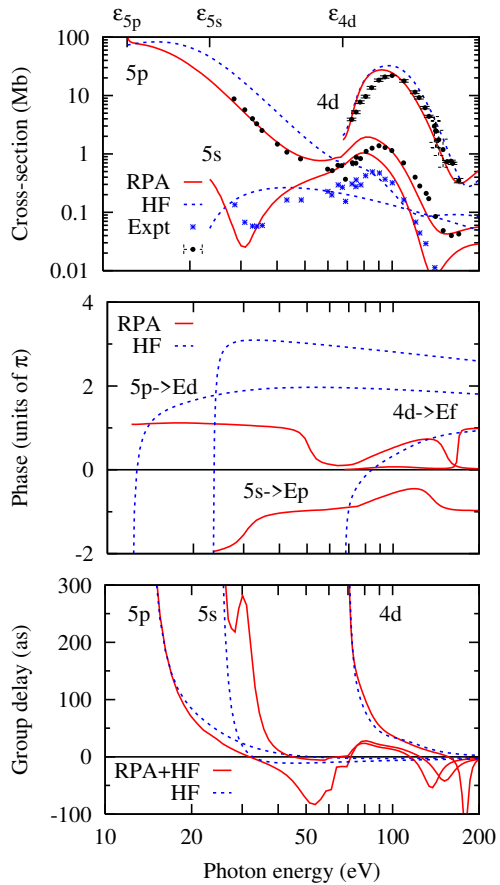}
\caption{(Color online) Top: the partial photoionization
cross-sections of Xe. The HF and RPA calculations are shown by the
dashed (blue) and solid (red for $5s, 5p$ and green for $4d$) lines,
respectively.
The experimental data from \protect\citet{PhysRevA.39.3902} and
\protect\citet{PhysRevA.30.812} are shown with asterisks for $5s$ and
error bars for $5p$ and $4d$.
Middle: elastic scattering phases in the field of the Ar$^+$ ion for
the $5s\to Ep$ and the dominant $5p\to Ed$ and $4d\to Ef$ channels
(dotted blue line) and the RPA phases (solid red line).  Bottom: the
phase derivatives are converted to the units of the group delay.
\label{FigXe}}
\end{figure}

The analogous set of data for the $5p$, $5s$ and $4d$ sub-shells of Xe is
presented in \Fref{FigXe}. On the top panel we compare the partial
photoionization cross-sections in the HF (dashed blue line) and RPA (
solid red line) approximations with the experimental data
\cite{PhysRevA.39.3902,PhysRevA.30.812} which are shown with the blue
asterisks for $5s$ and error bars for $5p$ and $4d$. The experimental
ionization thresholds $\rm \epsilon_{5p_{3/2}} = 12.13$~eV ,
$\epsilon_{2s}=23.40$~eV \protect\cite{NIST-ASD} and
$\epsilon_{4d_{5/2}} = 67.50$~eV \protect\cite{PhysRevA.40.5052} are
indicated on the upper boundary of the panel.

Below the $4d$ ionization threshold, the $5s$ and $5p$ cross-sections
in Xe behave similarly to to the $4s$ and $4p$ sub-shells in Kr (top panel
of \Fref{FigKr}). However, above this threshold, the $4d$ sub-shell goes
through a very steep shape resonance, sometimes even called a ``giant
resonance'' . This resonance is then turns into a Cooper minimum. By
strong inter-shell interaction, this behavior is replicated in the
$5p$ and $5s$ partial photoionization cross-sections which are well
reproduced by the RPA calculation. Accordingly, the corresponding RPA
phases displays steep $\pi$ jumps (middle panel) which are reflected
in the corresponding time delays (bottom panel).  In the case of the
$5s$ sub-shell, the RPA phase jump near the Cooper minimum mergers with
the Coulomb singularity and produces a very large, nearly 300~as time
delay at the photon energies below 30~eV. The $5p$ sub-shell shows a large
and negative time delay due to its Cooper minimum at around
50~eV. Both the $5s$ and $5p$ sub-shells display a large and negative time
delay near the local cross-section minima around 150~eV induced by the
correlation with the $4d$ sub-shell. The time delay in the $4d$ sub-shell is
driven from the strongly positive due to the Coulomb singularity at
low photon energies to a large negative jump near the Cooper minimum
at about 180~eV. At larger energies, the cross-sections are rather
structureless and there is no significant time delay variations.

\begin{figure}[h] \vs{0.45\cw} \epsfxsize=0.8\cw
\epsffile{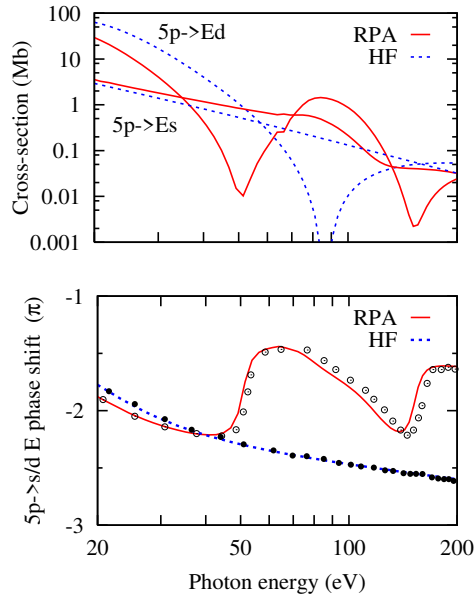}
\caption{(Color online) Top: Partial photoionization cross-sections of
Xe in the $5p\to Ed$ and $5p\to Es$ channels in the RPA (solid red
line) and HF (dotted blue line) approximations. Bottom: Phase shift
between the partial $5p\to Ed$ and $5p\to Es$ waves. The present RPA
and HF calculations (solid red and blue dotted lines, respectively)
are compared with the RPA and HF calculations reported in
\protect\citet{PhysRevA.64.062501} (open and filled circles,
respectively).
\label{FigXe2}}
\end{figure}

A phase jump of $\pi$, smoothed by the interaction between the two
channels, has already been observed both theoretically and
experimentally by analyzing the anisotropy parameter in
photoionization of Xe $5p$ sub-shell \cite{PhysRevA.64.062501}. This
parameter contains the phase shift between the two photoionization
channels with $l=l_i\pm1$. In the case of $5p$ photoionization, these
are $5p\to Ed$ and $5p\to Es$ transitions. Their partial
photoionization cross-sections and the relative phase shift are
presented on the top and bottom panels of \Fref{FigXe2}.  On both
panels, we show the present RPA and HF calculations displayed with the
solid red and blue dotted lines, respectively. On the bottom panel, we
exhibit the RPA (open circles) and HF (filled circles) phase shifts
reported by \cite{PhysRevA.64.062501}.
 
On the top panel of \Fref{FigXe2} we observe a significant shift of
the Cooper minimum in the $5p\to Ed$ channel towards the lower
photon energies and appearance of the secondary minimum due to the
correlation with the $4d$ sub-shell. In the meantime, the inter-shell
correlation does not change the $5p\to Es$ partial photoionization
cross-section in such a dramatic way. Accordingly, on the bottom panel
of \Fref{FigXe2}, we see a strong variation of the RPA phase shift
with the two successive jumps near the Cooper minima of the
$5p\to Ed$ cross-section. In the meantime, the HF calculation returns
quite a smooth and monotonous phase shift. Agreement between the two
sets of calculations, the present and the one reported by
\cite{PhysRevA.64.062501}, is rather good. A small shift between the
present calculation and the reference one is most likely due to
scanning and digitizing  the analog data of Fig.~3 in
Ref.~\cite{PhysRevA.64.062501}.

\section{Conclusion}
\label{Sec3}

In the present work, we perform a systematic study of the
photoemission time delay from the valence shells of noble gas atoms in
sequence from Ne to Xe. We cover the photon energy range from the
ionization threshold to 200~eV. We test the accuracy of our
calculation by making comparison with available partial
photoionization cross-sections. We derive the complex phase of the
photoionization amplitude in the non-relativistic HF and RPA
calculations and convert it to the photoelectron group delay by taking
the energy derivative.

The time delay results display a very diverse landscape due to an
interplay of three major factors. The first two are the logarithmic
Coulomb singularity and the Levinson theorem which drive the
photoelectron scattering phase in the field of the singly charged
ion. The third factor is the phase jump of $\pi$ near the Cooper
minimum which is smoothed by the inter-shell interaction. The two
former factors are revealed in the HF calculations whereas the third
one is most vividly reflected in the RPA calculations. Experimentally,
photoionization measurements near the Cooper minima may be
challenging but it is the area where the time delay effects are
expected to be largest.

These time delay results are compared with experimental data derived
from the attosecond streaking measurement \cite{M.Schultze06252010}
and the two-photon interferometric technique
\cite{PhysRevA.85.053424}. This comparison is inconclusive as the
difference between the theoretical and experimental results clearly
exceeds the reported error bars. We are fairly confident about the
accuracy of the present calculation which is tested by comparison of
the partial photoionization cross-sections with a large set of
independent experimental data and the angular asymmetry parameters as
in the case of Xe \cite{PhysRevA.64.062501}. It is hard to give a
numerical estimate on the accuracy of the group delay results. In
lighter atoms we expect it to be within 10\%. For Xe, it may be more
significant as suggested by larger difference between the calculated
and experimental cross-sections.  Even for this heaviest of the atoms
studied in the present work, the relativistic effects are not expected
to change considerably the complex phase \cite{PhysRevA.64.062501} and
hence the associated group delay.  It is therefore an open question
why the time delay results cannot be verified experimentally even
after the corresponding CLC or CC corrections are made. Such a
verification would be a very welcoming development both for the
attosecond time delay measuring techniques and the complete theory of
atomic photoionization. 

\section*{Acknowledgment}

The author is very thankful to Marcus Dahlstr\"om for communicating
his original time delay results \cite{PhysRevA.86.061402} in numerical
form and for performing an additional set of calculations
\cite{Marcus2013} which can be compared directly with the present
work.  The author  acknowledges Thomas Carette who also sent the time
delay data reported in \cite{PhysRevA.87.023420} in numerical form and
provided a detailed explanation of the peculiarities of the MCHF
method.  The author wishes to thank Vladislav Yakovlev and Anne
L'Huillier for many useful and stimulating discussions and a critical
reading of the manuscript.  This work is supported by the Australian
Research Council in the form of the Discovery grant DP120101805.


\np

\end{document}